\documentclass{article}
\usepackage{booktabs}
\usepackage{cellspace}
\usepackage{spconf,amsmath,graphicx}
\usepackage{todonotes}
\usepackage{color}
\usepackage{listings}
\usepackage{graphicx}
\usepackage{multirow}
\usepackage{pifont}
\usepackage{lipsum}
\usepackage{adjustbox}

\newcommand{\red}[1]{\textcolor{red}{#1}}
\newcommand{\cmark}{\ding{51}}%
\newcommand{\xmark}{\ding{55}}%
\usepackage{subcaption}

\title{DISCOVERING MALICIOUS SIGNATURES IN SOFTWARE FROM STRUCTURAL INTERACTIONS}
\name{%
\begin{tabular}{@{}c@{}}
Chenzhong Yin$^{1,*}$ \qquad Hantang Zhang$^{1,}$\sthanks{Both authors contributed equally to this work} \qquad Mingxi Cheng$^{1}$ \\
Xiongye Xiao$^{1}$ \qquad Xinghe Chen$^{1}$ \qquad \textit{Xin Ren}$^{1}$ \qquad \textit{Paul Bogdan}$^{1}$
\end{tabular}}

\address{$^{1}$University of Southern California, Los Angeles, CA, USA \\}
\graphicspath{{image/}}

\begin{document}

\maketitle

\begin{abstract}
Malware represents a significant security concern in today's digital landscape, as it can destroy or disable operating systems, steal sensitive user information, and occupy valuable disk space. 
However, current malware detection methods, such as static-based and dynamic-based approaches, struggle to identify newly developed (``zero-day") malware and are limited by customized virtual machine (VM) environments. 
To overcome these limitations, we propose a novel malware detection approach that leverages deep learning, mathematical techniques, and network science. 
Our approach focuses on static and dynamic analysis and utilizes the Low-Level Virtual Machine (LLVM) to profile applications within a complex network. The generated network topologies are input into the GraphSAGE architecture to efficiently distinguish between benign and malicious software applications, with the operation names denoted as node features.
Importantly, the GraphSAGE models analyze the network's topological geometry to make predictions, enabling them to detect state-of-the-art malware and prevent potential damage during execution in a VM.
To evaluate our approach, we conduct a study on a dataset comprising source code from 24,376 applications, specifically written in C/C++, sourced directly from widely-recognized malware and various types of benign software. The results show a high detection performance with an Area Under the Receiver Operating Characteristic Curve (AUROC) of 99.85\%. Our approach marks a substantial improvement in malware detection, providing a notably more accurate and efficient solution when compared to current state-of-the-art malware detection methods.
\end{abstract}
\section{Introduction}
\label{sec:intro}

\begin{figure*}
   \centering
   \includegraphics[width=\textwidth]{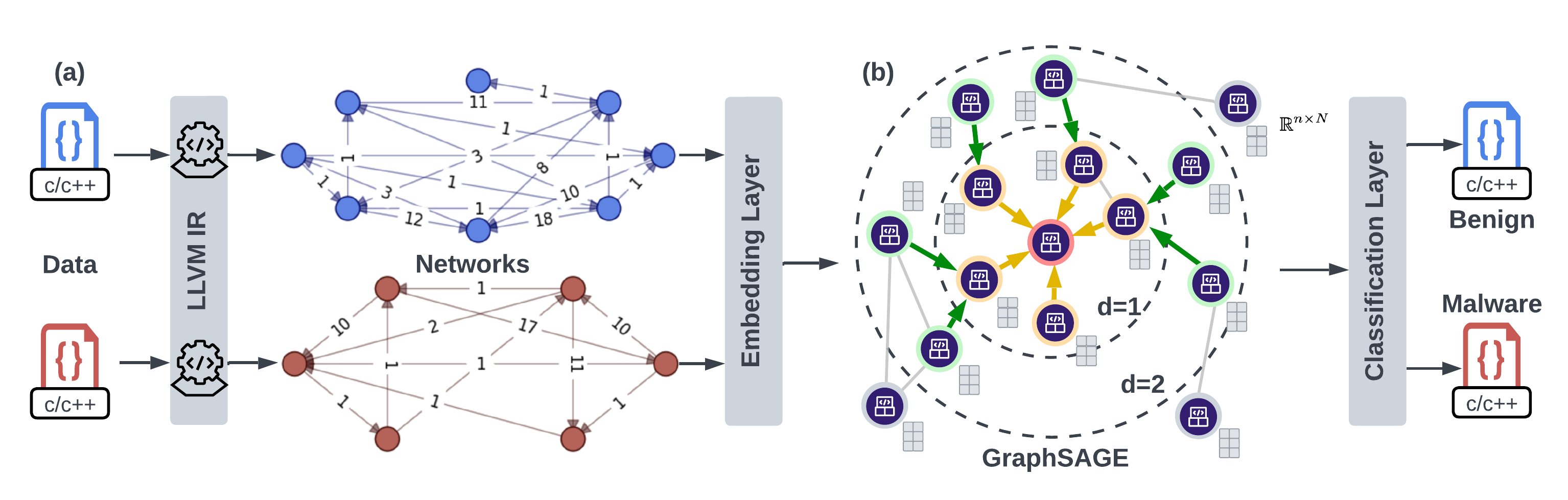}
    \caption{Overview of the Proposed Method: (a) C/C++ files are transformed into LLVM IR and converted into a weighted complex network by capturing operations from the source code. The generated networks are passed into an embedding layer to compute node features, resulting in an $n \times N$ matrix representation. (b) Leveraging these matrices, the GraphSAGE architecture extracts information from the neighbors of each node. Lastly, the classification layer predicts whether the file is malicious or benign.}
   \label{fig:cov_fig}
\end{figure*}

Malware refers to malicious software designed to cause damage to the Internet and computers. Malware could intrude into PC, collect user's personal information and sensitive data, and gain administrative privileges on a host, trigger havoc on PC's operating system, and lead to the loss of billions of dollars. For instance, malware-based cyber-attacks can target industrial control systems (ICS) to slow down industrial processes or destroy critical components, causing significant financial damages, safety, and life-threatening events~\cite{buchanan2022cyber}. 

To mitigate such catastrophic events, researchers have dedicated decades of work to malware detection. Mainstream solutions have emerged, namely static analysis and dynamic analysis. Static analysis involves scanning software binary byte-streams to generate signatures, like printable strings, n-grams, and instructions~\cite{mimura2022applying}. 
Kim et al. proposed a multimodal deep learning scheme to detect Android mobile malware by analyzing various factors such as strings, method APIs, permissions, components, and environment~\cite{kim2018multimodal}. Despite their high accuracy for specific malware types, static analysis methods struggle with ``zero-day" malware, where the signatures have changed due to updates~\cite{kumar2022zero}. 

On the other hand, dynamic analysis can handle the latest malware by monitoring their behavior during execution and regaining control of infected systems. Recently, researchers have applied deep learning models to dynamic analysis. Classic convolutional- (CNN) and recurrent- neural network (RNN) can learn features from the sequential data extracted from software~\cite{qiang2022efficient} \textcolor{black}{and image data~\cite{pelletier2019deep}}. For instance, Zhang et al.~\cite{zhang2020ransomware} designed a CNN-LSTM model to analyze features from each API call for detecting malicious files. 
\textcolor{black}{Furthermore, deep learning-based approaches can detect intrusion attempts~\cite{di2020experimental} and anomalies in network traffic~\cite{dong2021network}.}
Despite overcoming static analysis limitations, deep learning based dynamic approaches' inherent black-box nature hinders the interpretability of its feature attribution~\cite{yin2023anatomically}. 

To address the challenges posed by both static and dynamic analysis, this paper introduces a robust and mathematically grounded deep learning framework for malware detection called the Malware Graph Network (MGN). 
MGN involves adopting a Low-Level Virtual Machine (LLVM) intermediate representation (IR) compiler to transform suspected malware software into a complex weighted network, where nodes represent instructions, and edges indicate data and control dependencies among LLVM instructions. The weights associated with the network can represent data sizes or latency values. By utilizing this compiler approach, executing suspected applications on the virtual machine is avoided. 
This complex network representation allows capturing the spatiotemporal characteristics of software and interpreting the differences between benign and malicious files by revealing the intrinsic correlations between APIs and the software's instructions. To process these intricate network structures that have been extracted, we incorporate a GNN-based deep learning architecture called GraphSAGE into MGN for malware classification. This approach surpasses state-of-the-art baselines, such as N-gram Analysis~\cite{zhu2022n}, Grayscale Image Analysis~\cite{gibert2019using, almomani2022automated}, Operation Code Analysis~\cite{tang2022android}, and Byte File Analysis~\cite{chaganti2022deep, zhang2020ransomware}. \\
\vspace{-6mm}
\section{METHOD}
\vspace{-1mm}
To create a robust and interpretable malware detection strategy capable of effectively handling malware evolution and obfuscation, we present MGN, illustrated in Figure~\ref{fig:cov_fig}. MGN comprises two primary components: (1) utilizing a compiler approach built on LLVM to compile source code into a sophisticated network representation and (2) feeding the constructed networks into a GNN-based deep learning architecture for the purpose of malware detection. This two-part framework forms the foundation of our proposed approach for enhanced malware detection.

\subsection{Compiler Analysis and Network Modeling}

In this section, we leverage code compilation tools to transform source code into a graph representation. Specifically, we transform input applications into a corresponding dependency graph \cite{xiao2017load} through LLVM compiler and transparent by introducing IR as a common model for analysis, transformation, and synthesis. The LLVM is a compiler framework which makes program analysis lifelong. The resulting graph encompasses the structure of the graph itself, which includes nodes and the connections between them, as well as node features. For each graph, its node features are derived from the operation names present in the code, such as ``store", ``getelementptr", and ``load". 

We first run the applications with representative inputs to collect dynamic LLVM IR traces to resolve dynamic memory dependencies. Next, we analyze the traces to figure out whether source registers of the current instruction depend on destination registers of the previous instructions. If such dependencies exist, we add edges between these two nodes to represent dependencies. Then, these instructions are profiled to get the precise data sizes as edge weights in the graph. For example, a part of dynamic traces has been shown as follows: $\%3$ = sub $\%1$, $\%2$; $\%5$ = sub $\%3$, $\%4$. 
The second instruction has the register $\%3$, which depends on the destination register $\%3$ of the first instruction. 

Hence, we find a dependency between the last two instructions and an edge is inserted between these two nodes. 
After following these compiled steps, a suspected application is profiled into a complex network that will be analyzed in the following section to predict whether this application is benign or malicious. 
Figure~\ref{fig:cov_fig} shows the steps of dynamically profiling applications into graphs. 

After the transformation of the suspected files into graph representations, we encode the node features, represented by operation names, into one-hot vectors. 
Each graph's node features can be denoted by a matrix of dimensions $n \times N$, where $n$ signifies the number of distinct operation names present in a given graph and $N$ denotes a constant that represents the union of operation names across the entire dataset. Moreover, each graph is assigned a label indicating whether it represents malware or benign software.

\vspace{-3mm}
\subsection{GraphSAGE Neural Network Modeling}

In this study, we choose GraphSAGE to analyze the preprocessed features represented as complex networks. GraphSAGE can efficiently analyze large graph structures, where many software source codes often consist of numerous nodes~\cite{hamilton2017inductive}. 
Furthermore, GraphSAGE employs inductive learning and ensures excellent generalizability, which allows it to generate embeddings for previously unseen nodes. This character can be highly effective for classification tasks. Instead of relying on manually crafted features to describe malware, this property becomes invaluable in malware detection, given the frequent emergence of new malware variants and families.

Through an in-depth examination of GraphSAGE's structure and the graphs representing malware and benign software, we have devised our deep learning architecture. It comprises of the following components: an embedding layer, 6 GraphSAGE layers, 1 global pooling layer, and 1 output layer. In the input layer, we incorporate node features and edge indices, where the node features denote operation name for the instruction and the edge indices represent edge lists. 
The GraphSAGE employs the mean aggregator as its initial step, which calculates the mean of node features within the neighborhood. Subsequently, the number of neighbors sampled in each layer depends on the average degree of nodes. 

While GraphSAGE layer achieved the most exceptional performance in graph neural network model, the number of SAGE layers and the choice of activation functions have a significant impact on the final model's performance. This was evident in our ablation experiments.



\vspace{-3mm}
\section{EVALUATION}
\vspace{-3mm}

Following the same training and testing strategy with all the baselines, our model is trained and evaluated on 24,376 applications (12,815 malware and 11,561 benign software). 
All these applications are source code gathered from real-world programs. For malware representation, we select a diverse array of types, including Spyware, Botnet, Trojan, among others. As for benign software, we choose programs from gaming, system development, application software, mobile apps, and artificial intelligence to guarantee that our dataset comprehensively includes various domains. For every software, we identify their main execution functions as data points. These are then compiled using our toolset and transformed into a graph, serving as individual data entries.
The entire dataset is randomly split into $80\%$ as training set and the remaining $20\%$ is testing set. 
The model performances are assessed via the AUROC and accuracy (ACC). 

\begin{table}[h!]
\caption{Comparison of malware detection results using different hyperparameters and settings. The best results are highlighted in \textcolor{red}{red}.}
\vspace{-1mm}
\begin{adjustbox}{width=\columnwidth,center}
\begin{tabular}{c|c|c|c|c|c}
\hline \hline
{ \textbf{SAGE}}                 &  \textbf{EL} &  \textbf{LRelu} &  \textbf{ACC}   &  \textbf{AUROC}    &  \textbf{F1-score}      \\
\hline
\multirow{3}{*}{4layers} &  \xmark          & \xmark       & 96.29\% & 99.38\% & 96.81\%\\
                         & \cmark           & \cmark       & 97.45\% & 99.72\% & 97.76\%\\
                         & \cmark           & \xmark       & 97.17\% & 99.57\% & 97.54\%\\
                         \hline
\multirow{3}{*}{6layers} &  \xmark          & \xmark       & 95.79\% & 99.33\% & 96.08\%\\
                         & \cmark           & \cmark       & \red{98.55\%} & \red{99.85\%} & \red{98.72\%}\\
                         & \cmark           & \xmark       & 97.45\% & 99.73\% & 97.78\%\\
                         \hline
\multirow{3}{*}{8layers} & \xmark          & \xmark        & 96.51\% & 99.61\% & 96.96\%\\
                         & \cmark           & \cmark       & 98.18\% & 99.79\% & 98.40\%\\
                         & \cmark           & \xmark       & 97.45\% & 99.59\% & 97.78\%\\

                         \hline
\multirow{3}{*}{10layers} & \xmark         &  \xmark      & 95.65\% & 99.37\% & 95.95\%\\
                         & \cmark           &  \cmark     & 96.03\% & 99.46\% & 96.30\%\\
                         & \cmark           & \xmark      & 95.82\% & 99.40\% & 96.11\%\\

                         \hline \hline
                        
\end{tabular}
\end{adjustbox} 
\label{tab:1}
\end{table}



\vspace{-5mm}
\subsection{Ablation study}
To gain a comprehensive understanding of the significance of different components in our malware detection model and to determine the most effective architecture, we performed a systematic ablation study, varying hyperparameters and settings. Table~\ref{tab:1} presents a summary of our findings. In the table, ``EL" and ``LReLU" refer to the embedding layer and Leaky-ReLU activation function, respectively. The presence of a \cmark \ indicates that the embedding layer or Leaky-ReLU was utilized, while a \xmark \ denotes their absence (for activation function, we use ReLU instead). The first column represents the different numbers of SAGE layers chosen for the ablation study. 


In this study, we introduce an fully-connected embedding layer before the GNN architecture, which plays an essential role in capturing initial feature representations. The absence of this embedding layer results in a performance drop of approximately $2.76\%$. Furthermore, we incorporate 6 SAGE layers, each connected to a hidden layer with 128 hidden units. We choose Leaky-ReLU as the activation function for each SAGE layer.

\subsection{Experiment results}

\begin{figure}[h]
    \centering
    \includegraphics[width=0.5\textwidth]{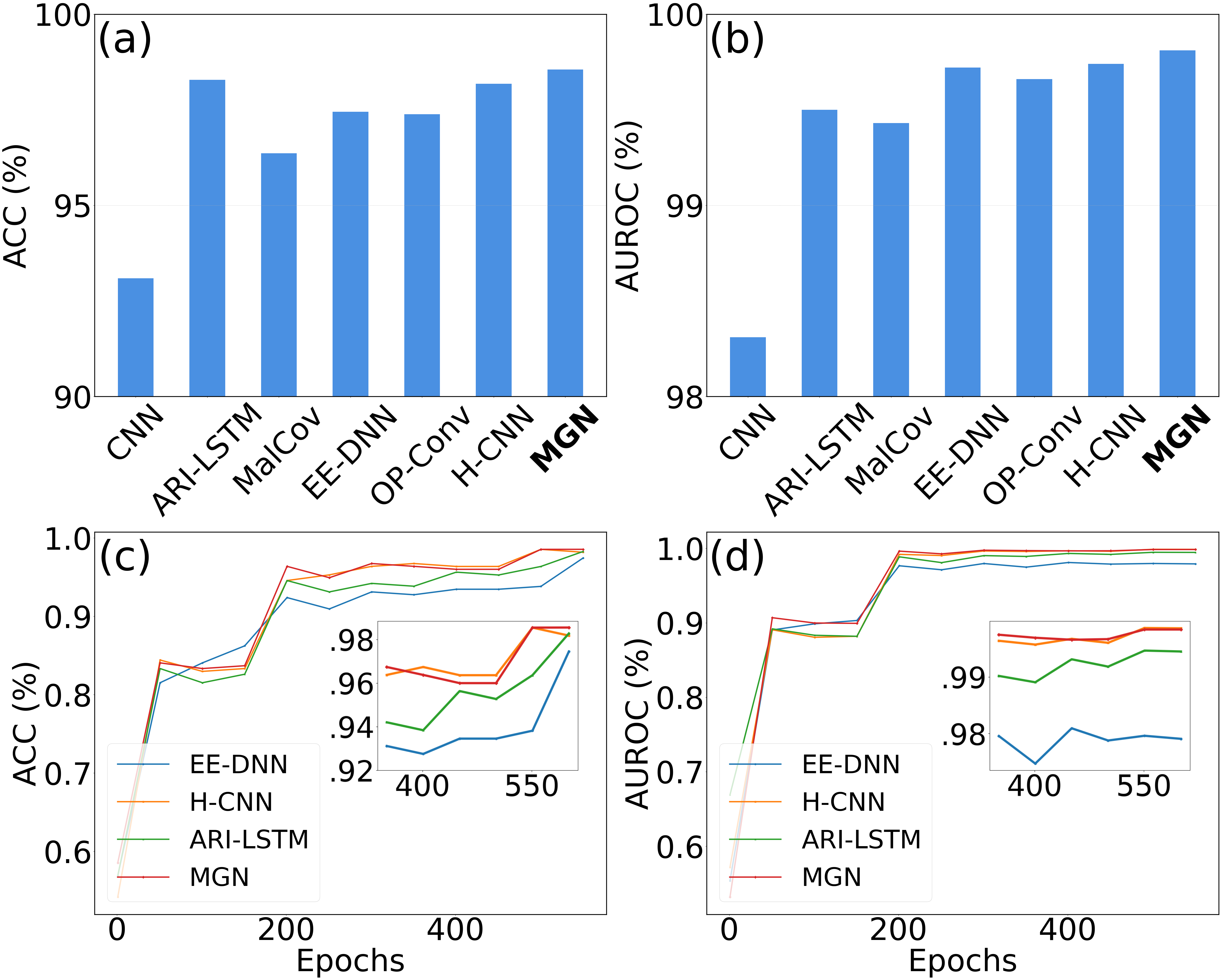}
    \caption{Performance comparison of MGN against baselines on the test set, showing (a) ACC and (b) AUROC; (c) ACC and (d) AUROC comparisons of MGN against baselines at different epochs.}
    \label{fig:fig3}
\end{figure}

The comparison results for malware detection are presented in Table~\ref{tab:tab2} and Fig.~\ref{fig:fig3}. In this section, we select state-of-the-art deep learning-based malware detection models as our baselines, as referenced in previous works~\cite{gibert2019using, agrawal2019attention, gibert2018end, zhang2020ransomware, gibert2019hierarchical, chaganti2022deep}. All these models are trained and tested on our malware dataset to ensure a fair comparison.

As indicated in Table~\ref{tab:tab2}, our MGN demonstrates remarkable performance in classifying malware and outperforms all other models in terms of both ACC and AUROC. To provide a visual representation of these comparisons, Fig.~\ref{fig:fig3} panels (a) and (b) display quantitative results for ACC and AUROC, respectively. Panels (c) and (d) respectively present the learning curve comparisons between MGN and selected baselines (specifically, ARI-LSTM, EE-DNN, and H-CNN, which utilize N-gram, Bytes, and Opcode as features, respectively).
Hence, our findings suggest that network-based features used in MGN can significantly enhance the performance of distinguishing between malware and benign software within the context of deep learning. For comprehensive information, please refer to Table~\ref{tab:tab2}.

Table~\ref{tab:tab3} presents a detailed breakdown of various malware types, along with their respective performance metrics obtained through our method and ARI-LSTM~\cite{agrawal2019attention}. We choose to include ARI-LSTM in this table because of its notable accuracy, as well as the similarity of its features to our node-based features.

\begin{table}[h!]
\caption{Comparison of malware detection results (ACC and AUROC) across different models. The best two results are highlighted in \textcolor{red}{red} and \textcolor{black}{blue}, respectively.}
\vspace{-1mm}
\begin{adjustbox}{width=\columnwidth,center}
\begin{tabular}{c|c|c|c}
\hline \hline
{ \textbf{Baselines}}                 &  \textbf{Features} &  \textbf{ACC} &          \textbf{AUROC}         \\
\hline

    CNN~\cite{gibert2019using} & Image & 93.09\% & 98.31\% \\
    \hline
    ARI-LSTM~\cite{agrawal2019attention} & N-gram  & \textcolor{black}{98.28\%} & 99.5\% \\
    \hline
    Bi-GRU-CNN~\cite{chaganti2022deep} & Bytes  & 96.36\% & 99.43\% \\
    \hline
    EE-DNN~\cite{gibert2018end} & Bytes  & 97.45\% & 99.72\% \\
    \hline
    SA-CNN~\cite{zhang2020ransomware}  & Opcode  & 97.38\% & 99.66\% \\
    \hline
    H-CNN~\cite{gibert2019hierarchical} & Opcode  & 98.18\% & \textcolor{black}{99.74\%} \\
    \hline
    MGN & Networks  & \textcolor{red}{98.55\%} & \textcolor{red}{99.85\%} \\
                         \hline \hline
                        
\end{tabular}
\end{adjustbox} 
\label{tab:tab2}
\end{table}

\begin{table}[h!]
\caption{Classification accuracy comparison between our model (MGN) and the state-of-the-art model (ARI-LSTM) across various malware types. The best results are highlighted in \textcolor{red}{red}.}
\vspace{-1mm}
\begin{adjustbox}{width=\columnwidth,center}
\begin{tabular}{c|c|c|c|c}

        \hline \hline
        \textbf{Malware} & \textbf{Samples} & \textbf{Text Samples} &\textbf{MGN} & \textbf{ARI-LSTM} \\
        \hline
        Spyware & 4757 & 950 & \textcolor{red}{98.32\%} & 98.11\%\\
        \hline
        Botnet & 1548 & 300 & \textcolor{red}{98.67\%} & 97.00\% \\
        \hline
        Trojan & 4645 & 900 & \textcolor{red}{99.11\%} & 98.89\% \\
        \hline
        Rootkit & 3048 & 600 & \textcolor{red}{98.33\%} & 97.67\% \\
        \hline
        Trojan-Backdoor & 3097 & 600 & \textcolor{red}{98.50\%} & 97.83\% \\
        \hline
        Worm & 1548 & 300 & \textcolor{red}{100.00\%} & 99.00\% \\
        \hline
        Ransomware & 900 & 180 & \textcolor{red}{100.00\%} & \textcolor{red}{100.00\%} \\
        \hline
        Injection & 900 & 180 & \textcolor{red}{98.89\%} & 98.33\% \\
        \hline
        Mixed & 3933 & 750 & \textcolor{red}{100.00\%} & 99.60\% \\
        \hline \hline
\end{tabular}
\end{adjustbox}

\label{tab:tab3}
\end{table}


\begin{figure}[h!]
    \begin{minipage}[b]{1.0\linewidth}
      \centering
    \centerline{\includegraphics[width=8.8cm]{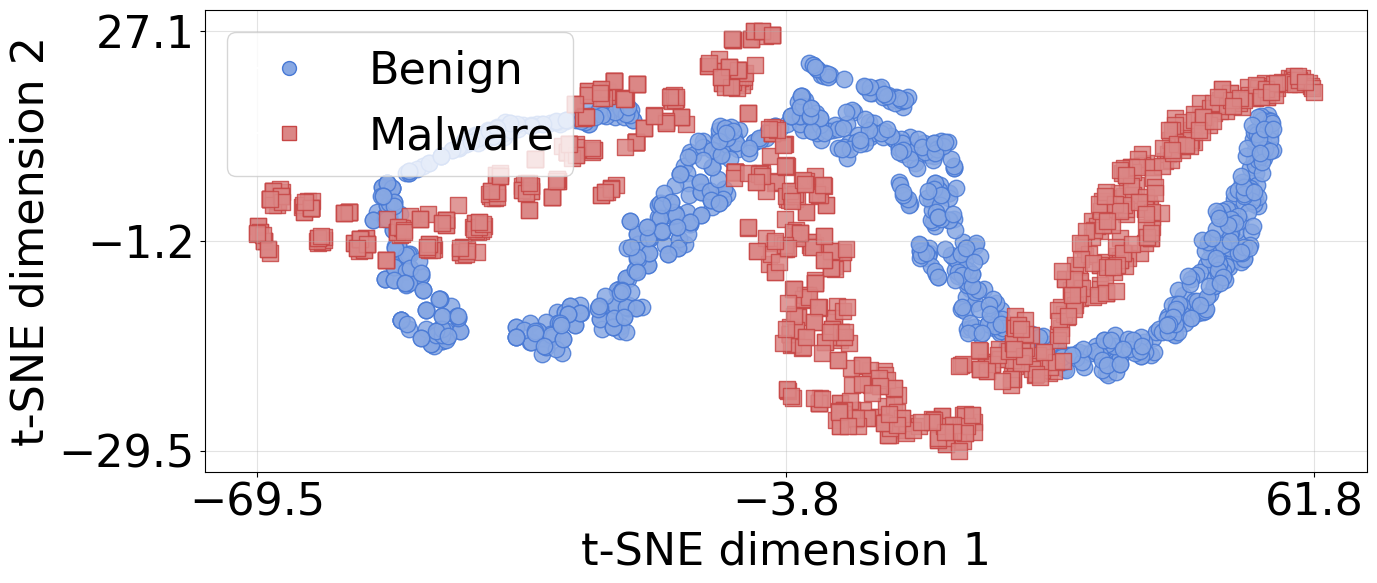}}
    \end{minipage}
    \caption{Two-dimensional t-SNE visualization of network topological features. Blue circles and red squares represent benign and malicious networks, respectively.}
    \label{fig.tSNE}
\end{figure}

\vspace{-3mm}
\subsection{Interpretability Analysis}

Figure~\ref{fig.tSNE} provides a clear two-dimensional t-SNE visualization, illustrating the clustering behavior of network topological features which includes the number of nodes and edges, average closeness-, average degree-, and average betweenness-centrality.These features are derived from the network representations of source files. We randomly selected around 900 instances from both the benign and malicious network classes for visual clarity.
This visualization effectively highlights the capability of our proposed methods in discerning the inherent topological distinctions between benign and malicious codes. Consequently, it enhances the interpretability of our approach.



\vspace{-4mm}
\section{CONCLUSION}
\vspace{-2mm}
In this paper, we propose an innovative deep learning framework known as MGN, designed for malware detection. MGN utilizes a graph-based representation and harnesses the power of the LLVM compiler to capture intricate software dependencies, resulting in the creation of a complex network. This network is then subjected to classification using a Graph Neural Network (GNN) architecture, significantly improving the precision of distinguishing between benign and malicious programs.
Through careful design of the model architecture, MGN demonstrates superior performance compared to state-of-the-art baselines, achieving higher accuracy and AUROC measurements. It is important to highlight that MGN's ability to transform software into the network's topology enhances its interpretability, setting it apart from other baseline approaches.

\bibliographystyle{IEEEbib}
\bibliography{strings}

\end{document}